\documentclass[12pt]{article}
\pdfoutput=1

\usepackage{scicite}
\usepackage[utf8]{inputenc}
\usepackage[T1]{fontenc}
\usepackage{color,soul}
\usepackage[super]{nth}
\usepackage{caption,booktabs,multirow,tabularx,ragged2e,dcolumn}
\usepackage{makecell}
\usepackage{graphicx}

\usepackage{times}

\newenvironment{sciabstract}{%
\begin{quote} \bf}
{\end{quote}}

\newcounter{lastnote}

\title{Metabolic light absorption, scattering and emission (MetaLASE) microscopy}

\date{}

\begin{document} 


\baselineskip24pt


\maketitle

\noindent \author{
Brendon S. Restall,$^{1^\dagger}$ Nathaniel J.M. Haven,$^{1^\dagger}$ Matthew T. Martell, $^{1^\dagger}$ Brendyn D. Cikaluk, $^1$ Saymon Tejay,$^3$ Benjamin A. Adam,$^2$ Gopinath Stendra ,$^3$ Xingyu Li,$^1$ Roger J. Zemp,$^{1^\ast\\}$

\noindent \footnotesize$^{1}$Department of Electrical and Computer Engineering, University of Alberta, 116 Street \& 85 Avenue, Edmonton, Alberta T6G 2R3, Canada\\
\footnotesize$^{2}$Department of Laboratory Medicine and Pathology, University of Alberta, 8440 - 112 Street, Edmonton, Alberta T6G 2B7, Canada\\
\footnotesize$^{3}$Department of Medicine, Faculty of Medicine \& Dentistry, University of Alberta, Edmonton, Alberta, Canada\\
\\
\footnotesize$^\ast$To whom correspondence should be addressed; E-mail:  rzemp@ualberta.ca \\
}


\begin{sciabstract}
 Optical imaging of metabolism can provide key information about health and disease progression in cells and tissues, however, current methods have lacked gold-standard information about histological structure. Conversely, histology and virtual histology methods have lacked metabolic contrast. Here we present a novel microscopy technology, Metabolic Light Absorption, Scattering and Emission (MetaLASE) microscopy, which rapidly provides a virtual histology and optical metabolic readout simultaneously. Photoacoustic remote sensing microscopy achieves nuclei contrast and hematoxylin contrast using ultraviolet absorption and an eosin contrast using the scattered ultraviolet light. The same ultraviolet source excites endogenous NADH, FAD and collagen autofluorescence allowing a measurable Optical Redox Ratios to see enhanced metabolism in areas of invasive carcinoma in breast and prostate tissues compared to benign regions. Benign chronic inflammation and glands also are seen to exhibit hypermetabolism. MetaLASE microscopy offers promise for future applications in intraoperative margin analysis, and in research applications where greater insights into metabolic activity should be correlated with cell and tissue types.
\end{sciabstract}



\section*{Introduction}

Cellular-level metabolism is critical in health and diseases such as cancer, fatty liver disease, autoimmune disorders, pancreatitis and more. Metabolic imaging has paved a way to better understand disease progression in live and fixed tissues. Optical metabolic imaging using the optical redox ratio (ORR), has proven to be a powerful means of tracking metabolism and differentiating tumor aggressiveness. Despite these advances, optical metabolic imaging methods have lacked co-registration with histology and when histology is needed, tissues are no longer viable. Metabolic status could still be of value in fixed tissues, but there are currently no methods for extracting both H\&E histology and metabolic status in the same tissue samples. To address these unmet needs, we introduce Metabolic Light Absorption, Scattering and Emission (MetaLASE) Microscopy, which offers realistic virtual histology of fixed or fresh tissues while simultaneously providing metabolic maps of the same tissue.

\subsection*{Metabolic and Virtual Histology Imaging Methods}

Non-optical metabolic imaging methods typically rely on metabolite imaging and glucose analogs such as FDG \cite{di2016potential}. The majority of optical metabolic microscopy methods have relied on autofluorescence emission from two metabolic electron carrier molecules of interest: NADH and FAD \cite{kolenc2019evaluating,georgakoudi2023label}. Due to different disease growth patterns there will be a variable glycolysis rate thereby generating additional nucleotides and increased NADH levels \cite{navas2021nad+}. Autofluorescence of NADH and FAD provides diagnostic information relating to cancer prognosis, aggression, and the presence of cancerous tissue \cite{kolenc2019evaluating,yaseen2013vivo,palmer2015detection}.

The optical redox ratio (ORR), defined as the ratio of FAD concentration to the sum of NADH and FAD concentrations \cite{chance1979oxidation,shiino1999three}, has emerged as a robust means of imaging metabolic state. A number of macroscopic optical metabolic imaging methods have been demonstrated \cite{shiino1999three,xu2010quantitative,xu2013redox,xu2016deep}. Microscopic metabolic imaging methods have included multi-photon excitation of endogenous chromophores \cite{skala2007vivo}, ultraviolet-excited autofluorescence microscopy \cite{AF_singleWave266nm,AF_singleWavemultiphoton}, and fluorescence lifetime microscopy (FLIM) \cite{skala2007vivo}. While macroscopic methods have lead to whole tissue evaluation, resolution is insufficient to investigate micro-scale features, achieve histologic details, or evaluate sub-cellular energetics. In contrast, microscopic methods have often been limited in field of view, precluding whole-tissue context and have still lacked histological realism for pathologist interpretation.

Histological information has been achieved with optical virtual histology methods, but current methods cannot yet achieve metabolic imaging simultaneously, despite some promising recent progress \cite{furtjes2023intraoperative}. 

The majority of label-free microscopy methods have lacked positive nuclei contrast. Haven et al., first demonstrated ultraviolet (UV) photoacoustic remote sensing (PARS) microscopy as means to provide positive nuclei contrast in reflection mode \cite{Haven1stUVPARS,haven2020reflective}. Photoacoustic remote sensing (PARS) microscopy is used to obtain absorption contrast in a label-free, non-contact approach, where a pulsed excitation source is used to generate reflectivity modulations in a co-focused interrogation beam, providing absorption contrast \cite{hajireza1stPARS,reza2018deep,haven2023investigating}. When a UV excitation source is used PARS provides cell nuclei specific contrast \cite{Haven1stUVPARS,haven2020reflective,Restall2021optomechanical,abbasi2019all,ecclestone2021single}. Additionally, cytoplasm contrast has been obtained simultaneously using back-scattered light, different excitation wavelengths, and frequency and time-domain decomposition of PARS signals \cite{restall2021virtual,kedarisetti2021f,bell2020reflection,pellegrino2022time}. PARS has further been combined with other optical imaging modalities such as optical coherence tomography and fluorescence microscopy \cite{martell2020multimodal,ecclestone2021three,martell2021fiber,restall2021multimodal}. Maximally realistic virtual H\&E histology was achieved using both UV absorption and UV scattering contrast as described in Haven et al. \cite{haven2021virtual}, and cycle-consistent generative adversarial network (CycleGAN)-based style transfer as demonstrated in Martell et al. \cite{martell2023deep}. Recently, we also validated these PARS virtual histology approaches in a pathologist reader study. For the task of identifying malignancy, pathologists viewing virtual histology images achieved sensitivity and specificity of 0.96 and 0.91 in breast tissues, and respectively 0.87 and 0.94 in prostate tissues \cite{martell2023deep}. Additionally, pathologists scored quality of virtual histology images to be superior to frozen sections, known to be prone to staining artifacts \cite{martell2023deep}. However, PARS virtual histology has not yet achieved metabolic contrast. 

\subsection*{MetaLASE Microscopy}

\begin{figure}[!htbp]
	\centering
\includegraphics[width = \linewidth]{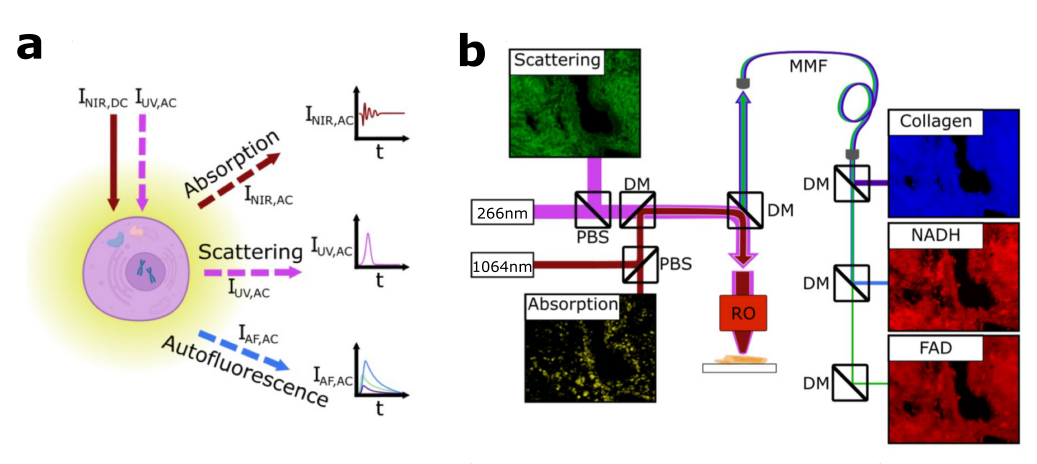}
	\caption{Simplified diagrams of the MetaLASE: \textbf{a}) absorption, scattering, and autofluorescence emission contrast mechanisms, and \textbf{b}) imaging system. $I_{NIR,DC}$: incident CW NIR interrogation intensity; $I_{UV,AC}$: incident pulsed UV excitation intensity; $I_{NIR,AC}$: back-scattered absorption-induced modulated NIR interrogation intensity; $I_{UV,AC}$: back-scattered pulsed UV excitation intensity; $I_{AF,AC}$: autofluorescence emission; t: time. PBS: polarizing beamsplitter; DM: dichroic mirror; RO: reflective objective; MMF: multimode fiber.}
	\label{fig:SimplifiedSystemDiagram}
\end{figure}

Here, we introduce MetaLASE microscopy, which is a functional extension to UV-PARS that simultaneously achieves both virtual H\&E histology and metabolic contrast for the first time. By leveraging a single UV excitation source, autofluorescence signatures from endogenous chromophores can be obtained in addition to absorption and scattering contrast used in virtual histology, as shown in Fig. \ref{fig:SimplifiedSystemDiagram}a. This offers comprehensive, co-registered structural and functional information in a single scan of the sample. By using only PARS and UV scattering to generate virtual histology images (which use only 266 nm and 1064 nm wavelengths), all other colors are available as information (including metabolic information) above and beyond H\&E virtual histology.

A simplified system diagram is shown in Fig. \ref{fig:SimplifiedSystemDiagram}b. We utilize two light wavelengths: a pulsed 266 nm laser and 1064 nm continuous wave (CW) super luminescent diode (SLD). The two beams are combined and co-focused using a reflective objective. The backscattered 266 nm light is used as a virtual Eosin channel for virtual histology. The modulated component of the 1064 nm light is the PARS signal and represents optical absorption of the 266 nm excitation beam. It is used as a virtual Hematoxylin channel. Pulsed 266 nm excited autofluorescence light is split into 3 channels representing signal dominantly from collagen, NADH, and FAD as further illustrated in Supplementary Information Fig. 1.

We develop and test our MetaLASE microscopy platform using formalin-fixed paraffin embedded (FFPE) tissues from both human prostate and breast, as well as live cell imaging.

\section*{Results}

To demonstrate the ability of PARS to image cell nuclei in fresh thick tissue, we excised a murine liver, mounted it beneath a coverslip secured onto a 3D-printed frame, while the frame and tissue were then mounted on a voice-coil stage for imaging \cite{cikaluk2023rapid}. Fig. \ref{fig:PARSFreshTissue} shows PARS images of cell nuclei in this fresh thick tissue.

\begin{figure}[!htbp]
	\centering
\includegraphics[width = \linewidth]{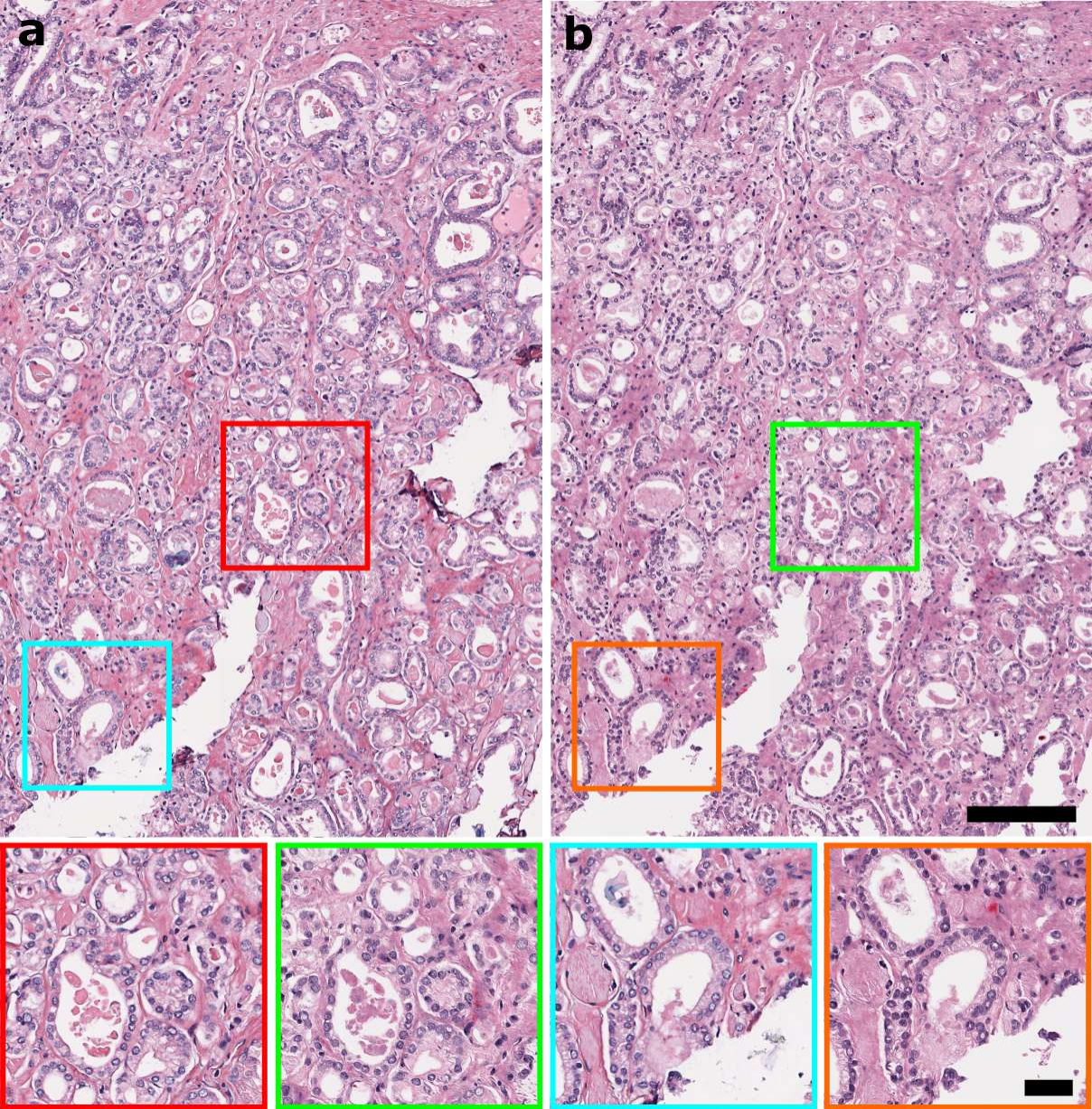}
	\caption{\textbf{a}) True H\&E-stained brightfield histology image of a sectioned radical prostatectomy specimen. \textbf{b}) Corresponding virtual histology image taken with our MetaLASE system. Scale bar: 200 $\mu$m. Colored square inset scale bars: 50 $\mu$m.}
	\label{fig:VirtualHEComp}
\end{figure}

To demonstrate the virtual histology capabilities of both UV absorption and scattering contrast with our MetaLASE microscope, we compared virtual H\&E histology images and true H\&E histology images of formalin-fixed thin sections of prostate tissue as shown in Fig. \ref{fig:VirtualHEComp}. The virtual H\&E histology image was formed as discussed in Martell et al. \cite{martellDeepLearning2022Preprint,martell2023deep}. Pathologist were accurately able to identify critical features of interest including invasive carcinoma, benign blood vessels, glands, and more.

To demonstrate both metabolic and virtual histology imaging, we used MetaLASE microscopy to image formalin-fixed thin sections since these could provide opportunities for histological validation. Example images of the optical redox ration (ORR), Virtual Histology, MetaLASE blended contrast image, and true histology image are show in Fig. \ref{fig:MetaLASEImage}. Zoom-in images outlined by the green box are visualized in Figs. \ref{fig:MetaLASEImage}e-m. The MetaLASE image shown in Figs. \ref{fig:MetaLASEImage}c and m are formed by superimposing absorption, scattering, collagen autofluorescence and optical redox ratios using yellow, green, blue, and pink colormaps, respectively. The respective NADH, FAD, ORR and collagen images are shown in Figs. \ref{fig:MetaLASEImage}e-h, respectively, while UV absorption and scattering images are shown in Figs. \ref{fig:MetaLASEImage}k and l. Virtual histology images show close agreement with true H\&E images in both Figs. \ref{fig:MetaLASEImage}b and d and in i and j, respectively. Beyond virtual H\&E and metabolic contrast, the collagen autofluorescence could serve as a label-free alternative to collagen stains such as trichrome stains.

\begin{figure}[!htbp]
	\centering
\includegraphics[width = 0.7\linewidth]{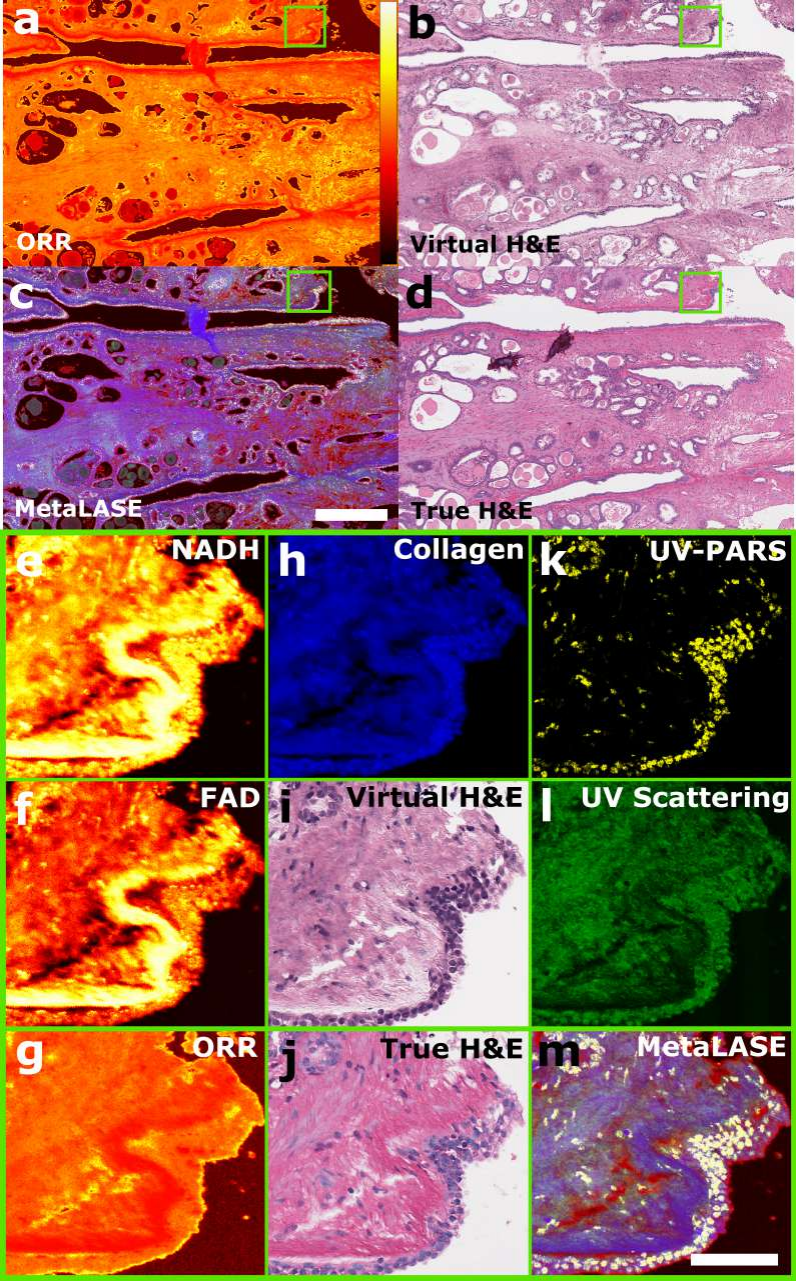}
	\caption{Radical prostatectomy tissue imaged with our MetaLASE system showing: \textbf{a}) ORR metabolic image; \textbf{b}) virtual histology image; \textbf{c}) combined MetaLASE image; \textbf{d}) corresponding H\&E-stained brightfield histology image. Scale bar: 500 $\mu$m. Colors in the MetaLASE image correspond to the following: blue: collagen; green: UV scattering; red: ORR; yellow: PARS. From the green square inset in \textbf{a}-\textbf{d}): \textbf{e},\textbf{f},\textbf{h},\textbf{k},\textbf{l}) different captured channel images with NADH, FAD, collagen, PARS, and UV scattering being shown, respectively; \textbf{g}) calculated ORR image from \textbf{e}) and \textbf{f}); \textbf{i}) virtual histology image obtained from \textbf{k}) and \textbf{l}); \textbf{m}) MetaLASE image obtained from \textbf{g}-\textbf{h}) and \textbf{k}-\textbf{l}). Scale bar: 100 $\mu$m.}
	\label{fig:MultiChannelExample}
\end{figure}

\begin{figure}[!htbp]
	\centering
\includegraphics[width = 0.8\linewidth]{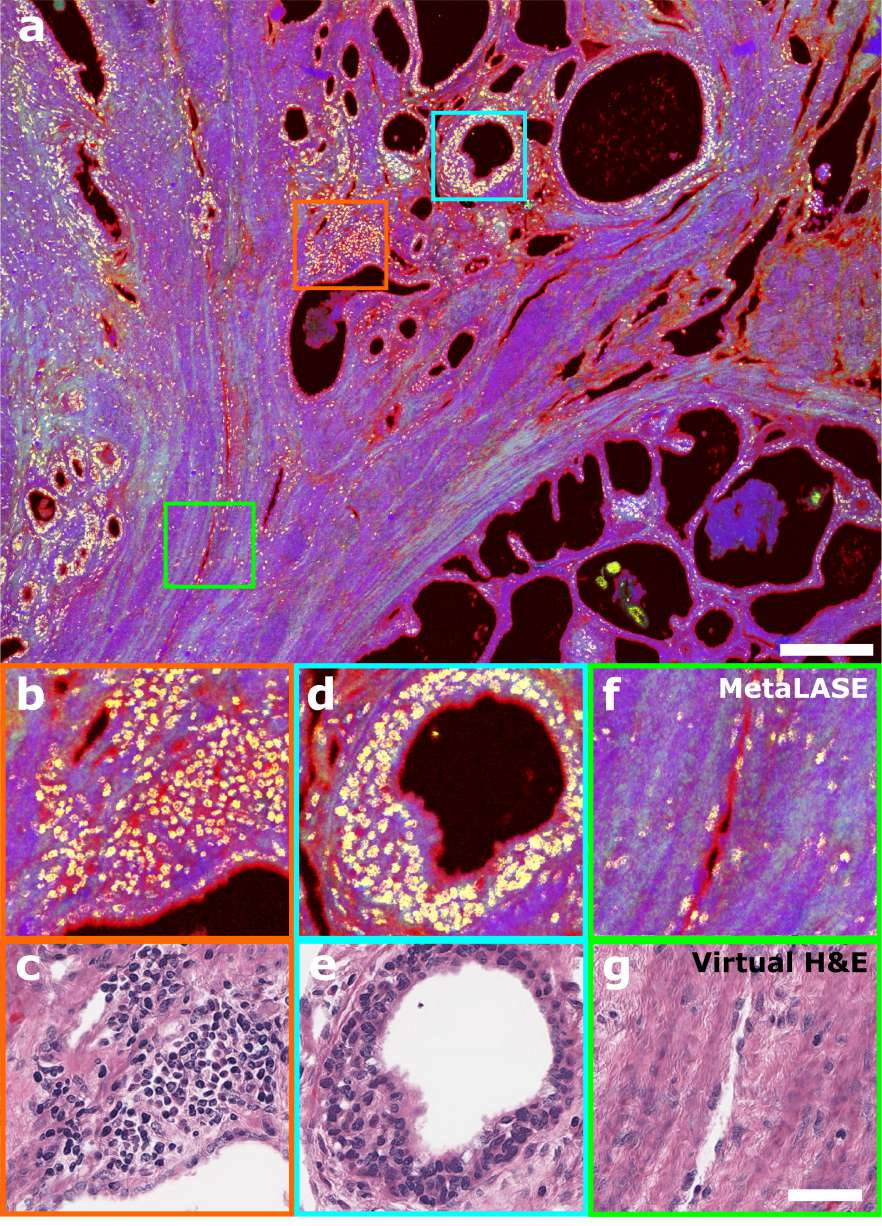}
	\caption{\textbf{a}) MetaLASE image of a sectioned radical prostatectomy specimen, scale bar: 200 $\mu$m. Colored square insets show zoom-ins of MetaLASE and corresponding virtual histology images in \textbf{b}/\textbf{c}, \textbf{d}/\textbf{e}, and \textbf{f}/\textbf{g}, respectively. Scale bar: 50 $\mu$m.}
	\label{fig:MetaLASEImage}
\end{figure}

Next, we investigated sources of metabolic contrast in tissues by imaging specimens from radical prostatectomies and from breast lumpectomies. Radical prostatectomy tissues containing Gleason 9 invasive carcinoma are imaged with coarse 50 micron-resolution over a wide-scale, as shown in the Supplementary Information Fig. 5.

Fig. \ref{fig:ORRVirtualHistology} demonstrates high metabolic activity in areas of benign inflammation in breast tissues. Fig. \ref{fig:ORRVirtualHistology} illustrates hypermetabolic regions around benign glands in prostate tissue. Since glands should be metabolically active, this is expected.

While hyper-intensity of ORR is not necessarily indicative of neoplasia, we found plenty of evidence that invasive carcinoma shows hyper-metabolism compared to benign regions. Fig. \ref{fig:ORRVirtualHistology} shows invasive carcinoma exhibiting higher average ORR in breast and prostate tissues compared to benign regions.

\begin{figure}[!htbp]
	\centering
\includegraphics[width = 0.7\linewidth]{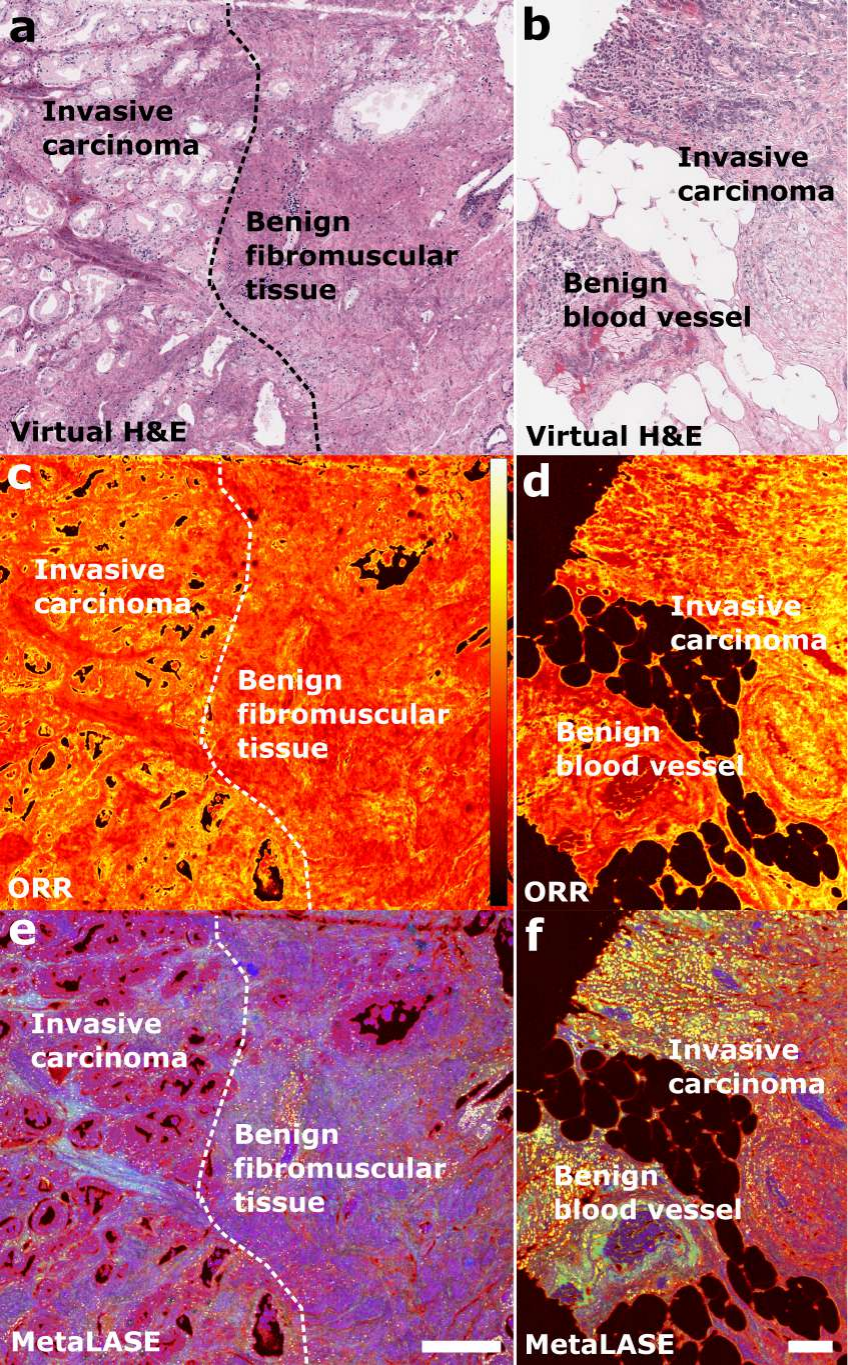}
	\caption{Virtual histology image of sectioned \textbf{a}) prostatectomy tissue, and \textbf{b}) lumpectomy tissue, with labeled regions of interest. \textbf{c}) and \textbf{d}) ORR maps for the corresponding virtual histology images shown in \textbf{a}) and \textbf{b}), respectively. \textbf{e}) and \textbf{f}) MetaLASE images for the corresponding virtual histology images shown in \textbf{a}) and \textbf{b}), respectively. \textbf{a},\textbf{c},\textbf{e}) Scale bar: 250 $\mu$m; \textbf{b},\textbf{d},\textbf{f}) scale bar: 125 $\mu$m.}
	\label{fig:ORRVirtualHistology}
\end{figure}

\begin{figure}[!htbp]
	\centering
\includegraphics[width = 0.8\linewidth]{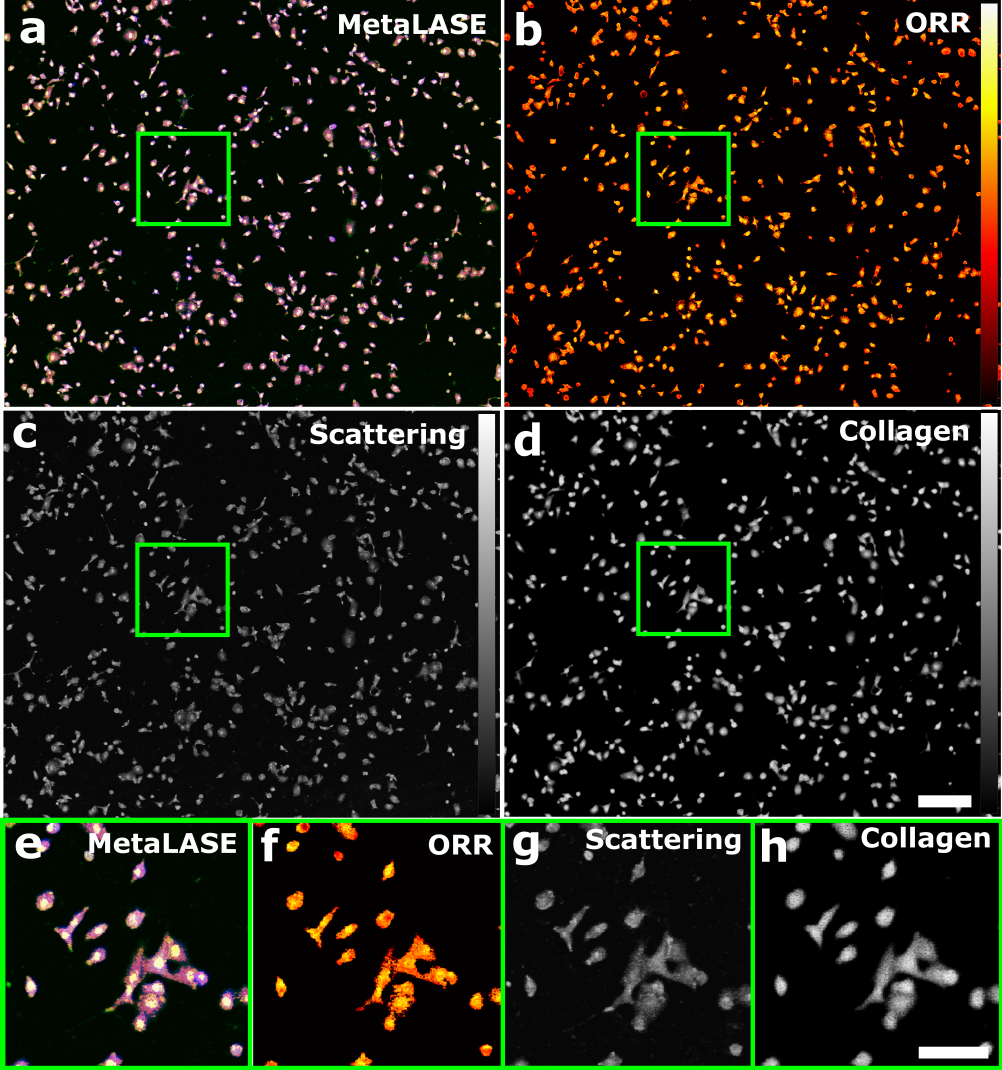}
	\caption{\textbf{a}) MetaLASE image of A549 lung carcinoma cells cultured on a glass slide in growth media, with accompanying ORR, UV scattering, and collagen contrast images in \textbf{b}-\textbf{d}), respectively, scale bar: 200 $\mu$m. \textbf{e}-\textbf{h}) Green square inset zoom-ins, scale bar: 50 $\mu$m.}
	\label{fig:LiveCells}
\end{figure}

\begin{figure}[!htbp]
	\centering
\includegraphics[width = \linewidth]{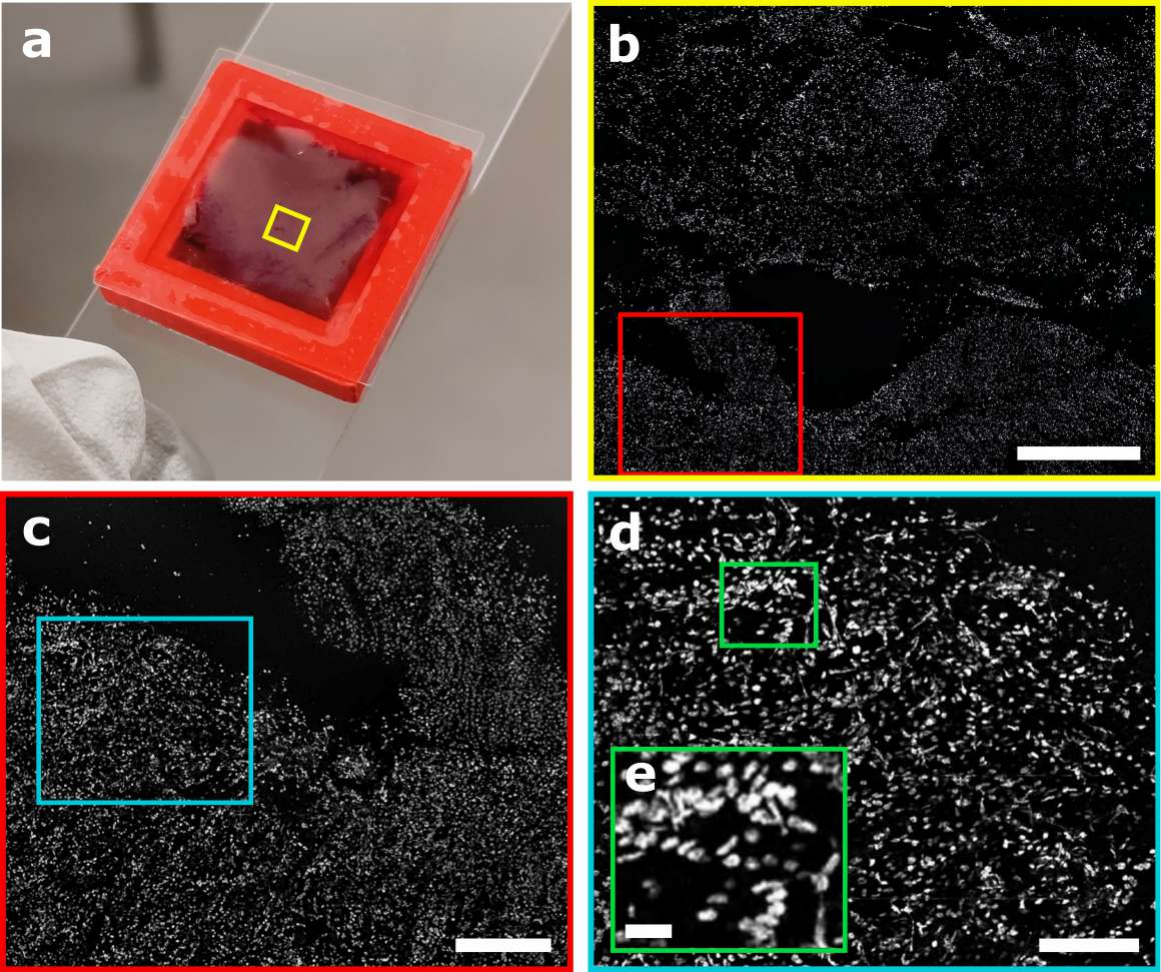}
	\caption{\textbf{a}) Image of fresh thick murine liver tissue compressed by a UV transparent coverslip. \textbf{b}) PARS microscopy image of yellow inset. Scale bar: 1 mm. \textbf{c}) Zoomed-in view of the red inset. Scale bar: 250 $\mu$m. \textbf{d}) Zoomed-in view of the cyan inset. Scale bar: 100 $\mu$m. \textbf{e}) Further magnification of green inset where indiviudal nuclei are clearly resolved. Scale bar: 15 $\mu$m.}
	\label{fig:PARSFreshTissue}
\end{figure}

To demonstrate live cell imaging, we imaged A549 lung carcinoma cells cultured on a glass slide in growth media. MetaLASE imaging results are shown in Fig. \ref{fig:LiveCells}. By successively zooming in, sub-cellular detail is observed where pink hotspots of high optical redox ratio are shown. These may correspond to mitochondria, the known location of NADH production. It is significant that we can see cellular energetics on a sub-cellular scale without the use of labels.

To demonstrate the ability of MetaLASE microscopy to image tumor specific metabolism, we imaged A549 lung carcinoma cells which were glutamine-fed or which were glutamine starved. Glutamine is an amino acid that is metabolized by tumors but not by normal non-neoplastic tissues \cite{hensley2013glutamine}. Tumors switch from glucose metabolism to this alternate energy source as outlined by many recent papers \cite{zhu2017metabolic,jiang2019starve,jin2023targeting,varone2014endogenous}. 24 hr starved cells had a large decrease in NADH concentration and the fed cells were observed to have a significantly greater mean autofluorescence level as seen in Fig. \ref{fig:Glutamine}, with $P < 0.001$ under a Welch's t-test of $n = 65$ image patches for each case. Each dataset was tested for normality by the Anderson-Darling test.

\section*{Discussion}

Our data demonstrates the ability to visualize cell nuclei with high contrast in thick fresh tissue, fixed tissue, and in live cells. This enables realistic virtual histology to be achieved with close concordance to true H\&E histology, as validated by pathologists. Additionally, our fast voice coil scanning system enables 1cm2 to be scanned with 400 nm resolution in 7minutes, rivaling or exceeding other microscopy platforms and even slide-scanner systems.

Moreover, MetaLASE demonstrates the unique ability to generate a rich information set above and beyond that provided in conventional histology, to visualize energetics of tissues, which may in turn offer clues as to aggressive phenotypes, response to therapies, and more. The fine resolution afforded by high-NA focusing of 266 nm UV light enables subcellular detail to be resolved, including visualizing hotspots in optical redox ratio with the cytoplasm, likely attributed to mitochondria. Thus, MetaLASE offers imaging scales from organelles to whole tissues.

Since MetaLASE microscopy is label free, it could be used to investigate samples repeatedly without permanent staining. It can also be used to image live cells and fresh tissues, whereas many stains require fixing, and this will alter and cease metabolism of the tissues.

Our experimental results demonstrating NADH autofluorescence changes due to glutamine metabolism could be of considerable interest in future work for discriminating aggressively malignant tumor tissue since glutamine is metabolized in tumors but not normal tissues.

One limitation of our work using formalin-fixed paraffin-embedded thin sections is the lack of tissue viability and recent work points out that metabolic measurements may be affected by the fixing process \cite{sanchez2023formalin}. Nevertheless, informative spatial variations in optical redox ratio and phenotype dependence have been observed by others and provides precedence to our imaging studies in fixed tissues \cite{li2020two}. 

In future work, MetaLASE could be used to investigate fresh biopsy specimens to produce virtual histology-like images at the point-of-care while also observing the ORR to estimate tissue metabolic activity for early prognosis. MetaLASE could further be used to image metabolic dynamics \cite{walsh2013optical,walsh2016optical,liu2018mapping} such as response to therapies \cite{walsh2014quantitative} and immune-cell activation \cite{walsh2021classification}.

Further analysis is required to see if we are able to observe any correlations between cell lines with a higher metabolism and specific phenotypes. We would predict that these would have a faster response to the metabolic rate changes and be another identifier of higher metabolism cells and tissue sections without the use of stains or labels to remove any human staining variability error \cite{stainError1,stainError2}. This could also include adding an electron transport chain (ETC) blocker like dichloroacetate (DCA) \cite{DCA_metabolismMichelakis} to the cell line and observe the metabolic rate change and reduced ORR concentration at different time intervals. This dynamic response measurement is important to quantify the ability of our system to quickly observe changes in metabolism and the effects on cell lines.

Future work would also include the utilization of deep learning algorithms to predict specific phenotypes from MetaLASE virtual histology \cite{jiang2021optical} and metabolic data. We could include data such as NADH and FAD, virtual pathological images, adjacent H\&E stained sections and each cancer’s phenotype profile to train a convolutional neural network (CNN). Currently, there is no other way to predict cancer aggression without time consuming and work intensive genetic testing and specific blood work, not available for all cancers. If possible this would have the potential to change critical diagnosis in borderline cancer cases, to provide earlier predictors of aggressive malignant transformation, early detection of therapy-resistant cancers and enable point-of-care pathology. 

\section*{Materials and methods}

\subsection*{Optical imaging system}

The MetaLASE microscopy system combines several modular subsystems to acquire multi-contrast data using a single ultraviolet excitation source. In the excitation path, 266 nm light is generated via second harmonic generation of a 532 nm nanosecond-pulsed source (SPFL-532-40, MKS) through a 4 x 4 x 10 mm cesium lithium borate (CLBO) crystal (Eksma Optics). The residual 532 nm beam is spectrally-separated from the 266 nm beam using a prism (PS863, Thorlabs) and is subsequently collected by a beam dump. The UV beam is then magnified through a variable beam expander (87-565, Edmund Optics) for achieving both collimation and filling the entrance pupil of the system objective. The polarization of the 266 nm beam is then rotated using a half-wave plate (HWP) (WPH05M-266, Thorlabs) and directed through a polarizing beam-splitter (PBS) (10SC16PC.22, Newport), quarter-wave plate (QWP) (WPQ05M-266, Thorlabs), and a 355 nm long pass dichroic mirror (DM) (Di01-R355-25x36, Semrock) for beam combination with the 1064 nm interrogation beam. Upon back-scattering, the UV beam will have an orthogonal polarization relative to the incident beam, and will thus be reflected by the PBS towards a 150 MHz amplified photodiode (PDA10A, Thorlabs) for 266 nm scattering contrast.

In the absorption detection path, the 1064 nm near-infrared (NIR) beam from a superluminescent diode (SLD) (SLD-1064-20-YY-350, Innolume) is fiber coupled into a circulator (HPBCIR-1060-H6-L-10-FA-SS, OF-Link) and passed through a zoom collimator (ZC618APC-C, Thorlabs), followed by a 900 nm longpass DM (DMLP900R, Thorlabs) and 355 nm long pass dichroic mirror (Di01-R355-25x36, Semrock) before being cofocused with the 266 nm beam onto the sample through a 0.5 NA reflective objective (LMM40X-UVV, Thorlabs). Back-scattered NIR light is then redirected via the circulator onto a 75 MHz balanced photodetector (PDB420C-AC, Thorlabs), providing absorption (PARS) contrast. 

After sample irradiance, the autofluorescence emission is collected by the system objective, and then transmitted through both the 355 nm and 900 nm longpass DMs and coupled into a multi-mode fiber (FG105UCA, Thorlabs). The fiber output is collimated (RC08FC-P01, Thorlabs) and directed into an enclosed subsystem for splitting of the tissue autofluorescence into distinct spectral bands corresponding to emission from collagen/elastin (352-405 nm), NADH (439-475 nm), and FAD (502-548 nm). The fluorescence emission is separated using a series of band-pass filters and dichroic mirrors corresponding to the collagen channel (84-093 - 377 nm, Edmund Optics) (FF414-Di01-25x36, Semrock), the NADH channel, (86-351 - 452 nm, Edmund Optics) (DMLP490R, Thorlabs), and the FAD channel (86-984 - 525 nm, Edmund Optics) (Di02-R635-25x36, Semrock). The spectral bands for each channel are then directed onto separate photomultiplier tubes (PMTs). For collagen, a PMT with a lower gain was utilized for collagen/elastin due to its significantly higher quantum yield and abundance (PMTSS, Thorlabs), and high sensitivity PMTs were utilized for NADH and FAD (H7422A-40, Hamamatsu Photonics).

The lateral spatial resolution of the various MetaLASE channels is primarily determined by the UV excitation spot size, previously measured as 390 nm \cite{haven2020reflective}.

\subsection*{Mechanical scanning system}

The scanning methodology in this work utilizes voice-coil stage scanning as demonstrated in Cikaluk et al. \cite{cikaluk2023rapid}. In brief, a linear drive voice-coil stage (X-DMQ12L-AE55D12, Zaber) rapidly oscillates along one transverse axis while a slow-axis stage (XMS-50S, Newport) traverses the orthogonal direction at a constant velocity, establishing a sinusoidal scanning trajectory over the sample surface. To trigger the laser, a function generator (DG1022Z, Rigol) generates pulses at a repetition rate ranging from 10 kHz - 2000 kHz, with the pulse repetition rate being chosen to obtain the desired spatial resolution for a oscillation frequency and scanning distance. The generated pulses are then used to trigger the laser and a digital delay generator (DG645, SRS) for external triggering of the data acquisition (DAQ) card for excitation event recognition, and for resetting of the system's analog peak detectors used in signal acquisition.

\subsection*{Data Acquisition}

In this work, eight data channels were acquired during imaging. For a precise reconstruction of the voice-coil position trajectory, the quadrature encoder channels from the stage must be recorded continuously over the duration of the scan. As such, a 100 MHz 14-bit digitizer card (CSE8389, GaGe Applied) was used to stream all data channels at 50 MS/s. The data acquisition channels include two quadrature encoder channels for absolute position reconstruction, a digital trigger signal for excitation event recognition, the PARS channel, the UV scattering channel, and three autofluorescence channels (collagen/elastin, NADH, FAD). The streaming rate of 50 MS/s was determined to be the maximum possible streaming rate for eight channel continuous streaming with the given acquisition system. Since the 20 ns sampling interval is too coarse to capture the rapidly developed optical signals, each signal is conditioned prior to digitization. The PARS signal at the output of the balanced photodetector was elongated via an inline 22 MHz low-pass filter (BLP-21.4+, Mini-Circuits), temporally broadening the waveform for sampling. Absorption-contrast data-points were extracted from the digitized waveform by integrating the PARS signal over a modulation window. To measure the magnitude of the back-scattered 266 nm pulses, a custom made peak detector\cite{snider2018toward} was used at the output of the UV channel photodiode prior to digitization. For the FAD and NADH autofluorescence channels, the output of the PMT (H7422PA-40) is amplified using a 300 MHz transimpedance amplifier (C11184, Hamamatsu), inverted using a 340 MHz operational amplifier (THS3001EVM, Texas Instruments) and subsequently passed through a peak detector for digitization. For the collagen/elastin channel, the output of the PMT (PMTSS) was amplified (MAX40662EVKIT\#, Analog Devices) prior to peak detection and digitization.

\subsection*{Image reconstruction}

To render data acquired from data acquisition channels as an image, the scan trajectory was reconstructed from AquadB encoder data using custom C++ OpenMP software for parallel read-in of encoder and optical signals, resolving stage positions from encoder state changes (fast axis) and sample timestamps (slow axis), and associating optical signal data with each position on the sinusoidal trajectory. Using Delaunay triangulation-based scattered data interpolation in a CPU-parallelized MATLAB script, this scan data was rendered onto a pixelated Cartesian grid using natural neighbor interpolation to form image arrays for each data channel. Total image formation required approximately 2.7 minutes per mm$^2$ using a 16-thread processor (i9-9900k, Intel) with 128 GB of RAM, with considerable opportunity for speedup of the interpolation component using GPU-based acceleration, and the scan trajectory reconstruction component using hardware-based (e.g. FPGA) implementations.

To generate virtual histology images, the PARS and UV scattering channels were utilized as virtual hematoxylin and eosin channels, respectively. To additionally achieve maximally-realistic stain-style, we used a cycle-consistent generative adversarial network, discussed in detail in Martell et al. \cite{martellDeepLearning2022Preprint, martell2022deep, martell2023deep}. In brief, PARS and UV scattering channels were assigned to red and green color channels in an RGB array, with the blue color channel remaining empty. Image color inversion was then performed to produce a background matching that of brightfield histology for network training. Over 15,000 unpaired pseudo-color virtual histology images and true H\&E histology images were subsequently used as inputs to a CycleGAN algorithm for training. Once trained, new pseudo-color inputs were transformed by the algorithm into maximally-realistic virtual H\&E images.

To generate a MetaLASE image, various optical contrast channels were combined into an RGB image. The red, blue, and green color channels were assigned to ORR, UV scattering, and collagen autofluorescence, respectively. Additionally, intrinsically sparse nucleic contrast was superimposed on the RGB image in a yellow-white colormap.

\subsection*{System calibration}

We calibrated the autofluorescence system to accurately measure the concentrations of NADH and FAD. We did so using a dilution curve of NADH and FAD and correlated the known concentrations to the measured voltage from the PMT, amplifier and subsequent DAQ card readings. NADH and FAD concentrations ranging between 5 $\mu$M to 500 $\mu$M were utilized. PMT voltage signal vs dilution concentration data was fit linearly with both fits having a calculated $R^2$ value over 0.93. See Supplementary Information Fig. 3 for the calibration plots.

\subsection*{Tissue acquisition \& preparation} 

Formalin-fixed breadloafed human lumpectomy and radical prostatectomy tissue specimens were obtained from breast and prostate cancer patients after pathology cases were closed and tissues were flagged for disposal as per approved ethics [HREBA (Cancer)/HREBA.CC- 20-0145]. Benign breast tissue specimens were obtained from a reduction mammoplasty procedures, where the tissue would have otherwise been discarded. Patient details pertaining to excised tissue specimens were kept anonymous from research staff and pathologists. Tissue specimens used in imaging were first paraffin embedded, then sectioned into 4 $\mu$m thin sections, followed by de-paraffination and re-hydration prior to imaging. De-paraffination of tissue sections was carried out by heat-adhering the tissues to the slides at 60°C for 1 hr, followed by 5 min washes in two changes of xylene, two changes of 100\% ethanol, 95\% ethanol, and deionized water.

\subsection*{Cell culture preparation} 

A549 cells were purchased and validated from ATCC (CCL-185), and cells were free from mycoplasma. Cells were cultured at 37 degrees in 9\% CO$_2$ in DMEM (Gibco \#11995-073; Glucose [25 mM], Glutamine [4 mM]) supplemented with 10\% FBS (Sigma-Aldrich \#F1051) and 1\% penstrep (Gibco \#15240062). For glutamine starvation experiments, cells were grown up to 60\% confluency in complete media, washed once with PBS then cultured in complete media or serum- and glutamine-free DMEM media (Gibco \#A1443001) for 24 and 48 hrs on glass slides prior to imaging.

\bibliography{scifile}

\bibliographystyle{ScienceAdvances}

\section*{Acknowledgements}

The authors are grateful for assistance with histology preparations and whole slide scanning from the Alberta Diabetes Institute HistoCore facilities and Shalawny Miller. The peak detection circuit used was originally designed by Logan Snider. The authors also thank Andrey Gorbunov for providing representative benign breast tissue samples and Lashan Peiris and Nadia Giannakopoulos for assisting with requisition of breast and prostate tissues. This work received funding from the Canadian Institutes of Health Research (PS 168936) and the Natural Sciences and Engineering Research Council of Canada (2018-05788). 

\section*{Author contributions statement}

Conceptualization: BSR, NJMH, MTM, BDC, GS, RJZ
Supervision: RJZ, GS, XL
Optical system design and experimental imaging: BSR, MTM, NJMH, BDC 
Hardware software development for scanning and image reconstruction: BSR, MTM, NJMH, BDC
Data Analysis: MTM, NJMH, RJZ
Clinical guidance and pathologist annotations: BAA
Manuscript Preparation: BSR, NJMH, MTM, BDC, RJZ
Manuscript editing and revisions: BSR, NJMH, MTM, BDC, ST, BAA, GS, XL, RJZ

\section*{Competing interests}

RJZ is a founder and shareholder of illumiSonics Inc., which, however, did not support this work. RJZ is founder, shareholder and director of CliniSonix Inc., which, however, did not support this work. RJZ is a scientific advisory board member of FUJIFILM VisualSonics, which, however, did not support this work.

\section*{Data availability}

The main data supporting the findings of this study are available within the paper and its Supplementary Information. Due to size considerations, raw data will be made available upon reasonable request.

\section*{Code availability}

The data processing and analysis described in Methods were implemented in MATLAB scripts which are available from the authors upon request.

\newpage

\nocite{broekgaarden2019tracking,reichert2023flavin,you2018intravital,ilie2019current,campbell2019non,wang2015high,mertz2011optical,HavenUVScatter, BCfast, ecclestone2022label,boktor2022virtual,boktor2023multi,tweel2023virtual,fereidouni2017microscopy,glaser2017light,xie2020diagnosing,tao2014assessment,cahill2020nonlinear,sun2020real,orringer2017rapid,hollon2020near,wong2017fast,imai2018high,cao2023label,kang2022deep,zhang2022high,soltani2022prostate,ragazzi2014fluorescence,perez2020basal,li2021biopsy}

\begin{table}[]
\includegraphics[width = \linewidth]{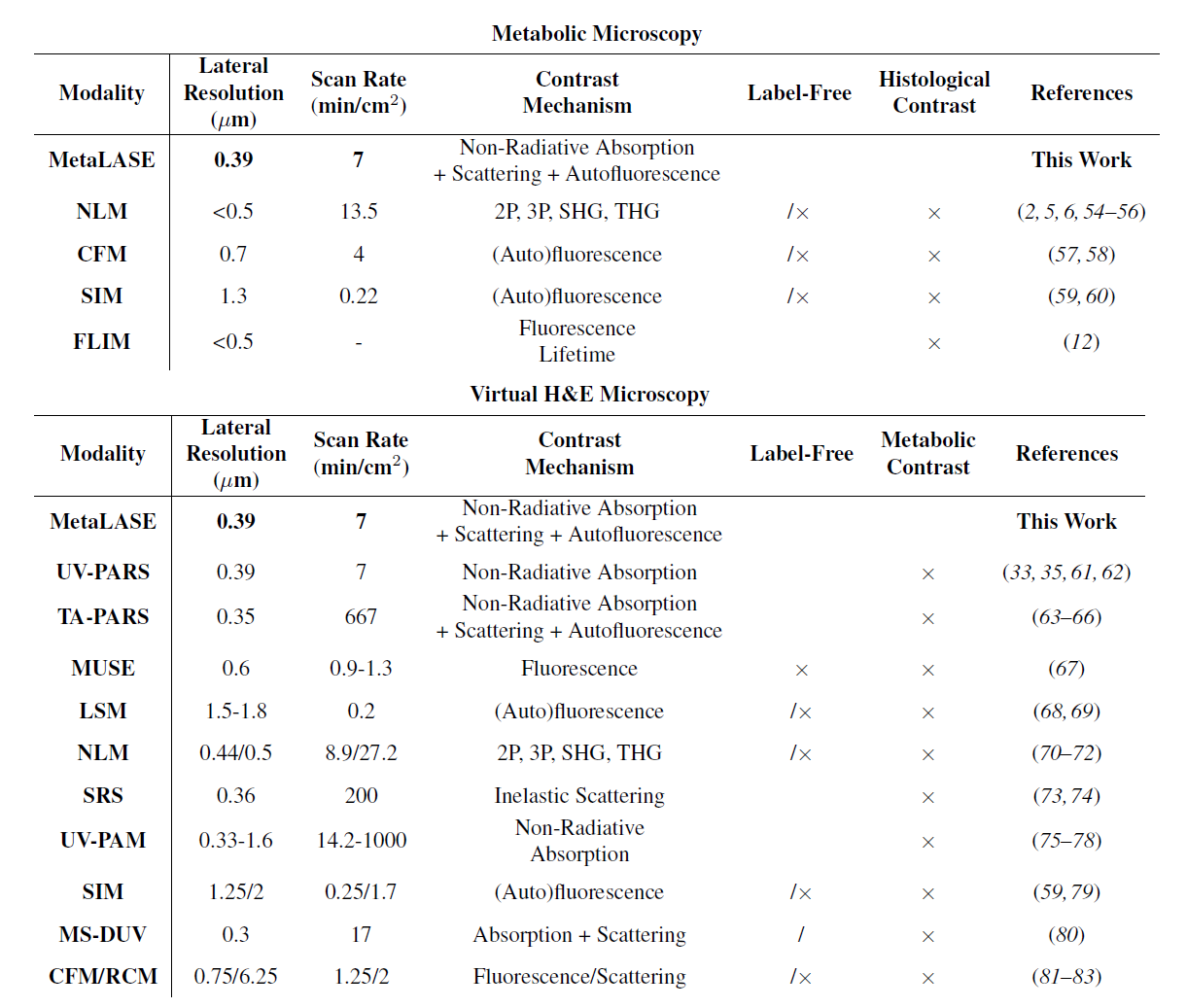}

\caption{Current metabolic and virtual histology modalities along with their resolution, scan speed, and contrast mechanism. Unlike many approaches in the literature, our virtual histology approach is label free and unlike many virtual H\&E microscopy implementations, our MetaLASE approach offers metabolic information. Note that specifications represent reported system implementations, but not necessarily fundamental limitations of each technology. 2P: two-photon; 3P: three-photon; SHG: second harmonic generation; THG: third harmonic generation; NLM: nonlinear microscopy; CFM: confocal fluorescence microscopy; SIM: structured illumination microscopy; FLIM: fluorescence lifetime imaging microscopy; TA-PARS: total absorption PARS; MUSE: microscopy with ultraviolet surface excitation; LSM: lightsheet microscopy; SRS: stimulated Raman scattering; PAM: photoacoustic microscopy; MS-DUV: multispectral deep UV; RCM: reflectance confocal microscopy.}
\label{table:1}
\end{table}

\newpage

\newpage

\begin{figure}[!htbp]
	\centering
\includegraphics[width = \linewidth]{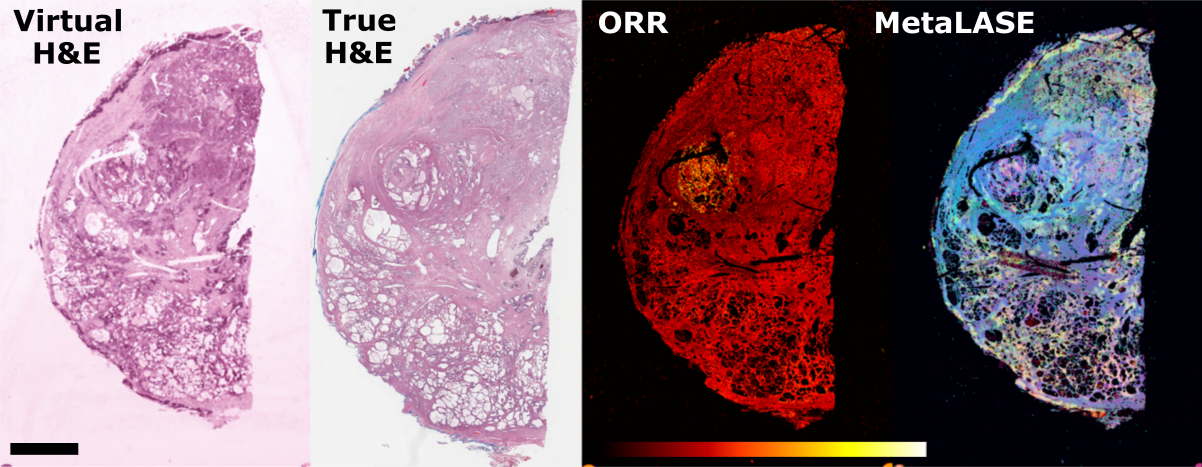}
	\caption{MetaLASE, optical redox ratio image, and virtual histology in a 3 cm long radical prostatectomy section with corresponding true H\&E-stained histology. Scale bar: 5 mm.}
	\label{fig:}
\end{figure}

\newpage

\begin{figure}[!htbp]
	\centering
\includegraphics[width = \linewidth]{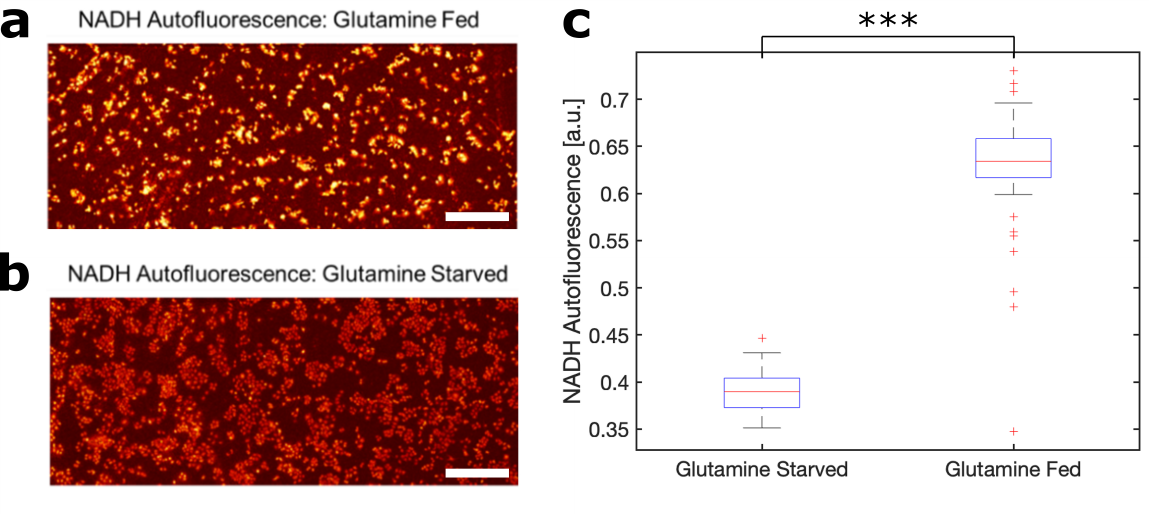}
	\caption{NADH autofluorescence images for A549 lung carcinoma cell culture under \textbf{a}) glutamine fed and \textbf{b}) 24 hr glutamine starved conditions. Scale bar: 100 $\mu$m. \textbf{c}) Box-and-whisker plots for mean NADH autofluorescence signal analyzed for $n = 65$ image patches under each condition.}
	\label{fig:Glutamine}
\end{figure}

\end{document}